\documentclass{emulateapj}
\def\mathbi#1{\textbf{\em #1}}
\usepackage{amsmath}
\usepackage{graphicx}
\shorttitle{Oscillations in solar waveguides}
\shortauthors{Luna-Cardozo, Verth, \& Erd\'{e}lyi}

\begin{document}
\title{Longitudinal oscillations in density stratified and expanding solar waveguides}

\author{M. Luna-Cardozo \altaffilmark{1,2},
          G. Verth \altaffilmark{3}, and
          R. Erd\'{e}lyi \altaffilmark{2} }
\email{mluna@iafe.uba.ar, gary.verth@northumbria.ac.uk, robertus@sheffield.ac.uk}
%
\altaffiltext{1}{Instituto de Astronom\'{i}a y F\'{i}sica del Espacio, CONICET-UBA, CC. 67, Suc. 28, 1428 Buenos Aires, Argentina.}
\altaffiltext{2}{Solar Physics and Space Plasma Research Centre (SP$^2$RC), University of Sheffield, Hicks Building, Hounsfield Road, Sheffield S3 7RH, UK.}
\altaffiltext{3}{School of Computing, Engineering and Information Sciences, Northumbria University, Newcastle Upon Tyne, NE1 8ST, UK.}

\begin{abstract}
Waves and oscillations can provide vital information about the
internal structure of waveguides they propagate in.
Here, we analytically investigate the effects of density and
   magnetic stratification on linear longitudinal magnetohydrodynamic (MHD)
   waves.  The focus of this paper is to study the
   eigenmodes of these oscillations. It is our specific aim
   to understand what happens to these MHD waves generated in flux tubes with non-constant
   (e.g., expanding or magnetic bottle) cross-sectional area and density variations.
The governing equation of the longitudinal mode is derived
   and solved analytically and numerically. In particular, the limit of the thin
   flux tube approximation is examined.
The general solution describing the slow longitudinal MHD waves
   in an expanding magnetic flux tube with constant density is
   found. Longitudinal MHD waves in density stratified loops with constant
   magnetic field are also analyzed.
From analytical solutions, the frequency ratio of the first overtone and fundamental mode is investigated in stratified waveguides. For small expansion, a linear dependence between the frequency ratio and the expansion factor is found.
From numerical calculations it was found that the frequency ratio strongly depends on the
   density profile chosen and, in general, the numerical results are in
   agreement with the analytical results.  The relevance
   of these results for solar magneto-seismology is discussed.
\end{abstract}

%

\section{Introduction}

High-resolution imaging and spectroscopic data from the
{\it Transitional Region and Coronal Explorer} ({\it TRACE}), {\it Solar
and Heliospheric Observatory} ({\it SOHO}), {\it Solar Terrestrial
Relations Observatories} ({\it STEREO}) and {\it Hinode} have revealed a variety
of wave modes in solar magnetic structures in the solar atmosphere
(see, e.g., Banerjee et al. 2007; Aschwanden 2009; Taroyan \&
Erd\'{e}lyi 2009; De Moortel 2009; Jess et al. 2009; Mathioudakis et al. 2011).
Observations of standing
slow magnetohydrodynamic (MHD) waves have been reported by different authors, using {\it SOHO}/SUMER
(Kliem et al. 2002; Wang et al. 2002, 2003, 2005), Yohkoh
(Mariska 2005, 2006) and recently, {\it Hinode}/EIS (Erd\'{e}lyi \& Taroyan 2008).
For an extensive review of the observation and modeling of 
standing slow magnetoacoustic waves see, e.g., Wang (2011).
Coronal seismology, originally suggested by Uchida (1970), Zaitsev \& Stepanov (1983), and Roberts et al. (1984), allows to obtain various physical
parameters (e.g., magnetic field and density scale height) through matching the MHD theory and waves observations in the corona.
The concept was proposed to be used in any magnetic structure of the Sun by Erd\'{e}lyi (2006) and labeled as solar magneto-seismology.
The topic was extensively reviewed with plenty of references by Andries et al. (2009) and Ruderman \& Erd\'{e}lyi (2009).

The theory of MHD wave propagation in magnetic structures in the
solar corona has been developed by modeling the magnetic structures as
homogenous cylindrical magnetic tubes embedded within a magnetic environment
(Rae \& Roberts 1982; Edwin \& Roberts 1983; Roberts et al. 1984).
Erd\'{e}lyi \& Fedun (2010) generalized these analytic efforts for fully
compressible twisted magnetic flux tubes.  Perhaps the simplest model
to study the oscillatory properties is a
cylindrical flux tube in pressure balance, without the complexities
of gravity, curvature, radiation, bulk motion, or heating. In this
case, such modeling leads to a profile with constant pressure,
density, and temperature along the tube.

However, more advanced equilibrium models have also been proposed with, e.g.,
dissipative effects and gravity (Mendoza-Brice\~{n}o et al. 2004), and it
has been found that the decay times of the standing slow modes are reduced
by 10\% - 20\% due to gravity when compared to the non-stratified loop models.
Mendoza-Brice\~{n}o \& Luna-Cardozo (2006) included radiative cooling and
heating on the study of longitudinal oscillations in hot, isothermal
coronal loops with constant coronal heating; it was found that the lack of
balance between cooling and heating does not affect the shape and decay
time of the oscillations.

Sigalotti et al. (2007, 2008) studied standing slow waves in hot coronal loops
finding that in order to achieve the same rate of damping time as detected in
the observations,
compressive viscosity has to be considered along with thermal conduction.
Non-isothermal profiles of longitudinal waves in hot coronal loops were
examined numerically, arriving at longer periods and shorter damping times
when the loop becomes more non-isothermal (Erd\'{e}lyi et al. 2008;
Luna-Cardozo et al. 2008).

However, the theory used in these previous studies assumed a constant magnetic
field along loops. Verth \& Erd\'{e}lyi (2008) investigated the combined effects of magnetic
and density stratification on transversal coronal loop oscillations.
It was found that even a relatively small coronal loop
expansion can have a significant and pronounced effect on the accuracy
of the plasma density scale height measurements derived from observations of
loop oscillations (see, e.g., Verth et al. 2008 for observational case study). Soon after, Ruderman et al. (2008) applied the theory and
found that the estimated coronal scale height is a monotonically decreasing
function of the tube expansion, while studying transverse oscillations
in a coronal loop with variable circular cross-sectional area and plasma
density in the longitudinal direction.

On average, the magnetic field strength is expected to decrease with
height above the photosphere, although it has been very difficult to
measure directly the variation of it in the corona.
However, Lin et al. (2004) had made some progress using spectropolarimetry,
and the results seem to confirm the decreasing with height of the magnetic
field strength.  It is expected that most loops should expand with height above
the photosphere, since the flux tube cross-sectional area and magnetic field
strength are inversely proportional. This expansion is defined here by
\begin{equation}
\Gamma= \frac{r_\mathrm{a}}{r_\mathrm{f}}, \label{eq01}
\end{equation}
where $r_\mathrm{a}$ is the radius at the apex and $r_\mathrm{f}$
is the radius at the footpoint (Klimchuk 2000).  The loop expansion has been
estimated for a number of loops, for example, Watko \& Klimchuk (2000)
reported mean values of 1.16 and 1.20 for nonflare and postflare EUV loops,
respectively, analyzing {\it TRACE} data. Klimchuk (2000) measured
a median value of $\Gamma \approx 1.30$ for soft X-ray loops using
Yohkoh data. However, potential and magnetic field extrapolation had given
larger loop expansions than the observed ones. DeForest (2007) suggested a possible 
explanation for this based on the fact that resolutions of images have not been 
sufficient to actually detect coronal loop expansion.
Regarding the chromosphere, it has been suggested theoretically that flux tubes 
must undergo significant expansion with height, the so-called magnetic canopy model, 
e.g., Gabriel (1976). However, to date there is little observational evidence to 
support this (Zhang \& Zhang 2000). More recently, using the Solar Optical Telescope (SOT) on board {\it Hinode}, for the first time Tsuneta et al. (2008) estimated an upper bound for chromospheric area expansion in the Sun's south polar region to be a factor of 345, giving a maximum expansion factor of approximately 19 for chromospheric flux tubes.

In this paper, the governing equation of the longitudinal MHD mode is derived and solved for
two representative cases modeling solar atmospheric flux tubes: an expanding magnetic flux tube with arbitrary longitudinal plasma density
and a density stratified flux tube with constant magnetic field.
We examine the governing wave equation within the limit of the thin flux tube
approximation. The slow mode is decoupled from the other MHD
modes, in a similar way as applied by D\'{i}az \& Roberts (2006), where the
slow mode was studied in density-structured coronal loops with constant
magnetic field. The purpose of this paper is to quantify the separate effects of the expansion factor $\Gamma$ and density stratification.
The shooting method is applied to find numerical solutions of the general
wave governing equation to compare them with analytical approximations.

\section{Magnetic field equilibrium configuration}
\label{potential}
The magnetic field equilibrium that decreases in strength with
height above the photosphere is modeled by an expanding flux tube with
rotational symmetry about the $z$-axis in cylindrical coordinates ($r,\theta,z$). Neglecting curvature along the tube axis we model an expanding tube with a straight central axis, i.e., a magnetic bottle. The tube ends are frozen in a dense photospheric plasma at $z=\pm L$,
and the flux tube has an arbitrary density depending on $z$.
This expanding magnetic field in equilibrium has two components
\begin{equation}
\mathbi{B} =B_r(r,z) \mathbi{e}_r +B_z(r,z) \mathbi{e}_z . \label{eq02}
\end{equation}

Following the derivation of Verth \& Erd\'{e}lyi (2008) for a potential field configuration, if small expansion is assumed it is
possible to obtain an explicit expression
for the perpendicular distance from the tube axis to a magnetic surface with footpoint distance from the axis defined by $r_f$, i.e.,
\begin{eqnarray}
r_0(z) \approx r_{\mathrm{f}} \left\{ 1+\frac{(1-\Gamma^2)}{\Gamma^2}
\frac{\left[\cosh \left(z/L \right)-\cosh(1)\right]}{1-\cosh (1)}
\right\}^{-1/2}. \label{eq03}
\end{eqnarray}
Note that in the thin tube approximation Equation (\ref{eq03}) is only a function of $z$. Similarly, 
the magnetic field components $B_r$ and $B_z$ at near the tube axis can also be described explicitly as a function of $z$,
\begin{equation}
B_r(z)\approx -B_{z,\mathrm{f}} \left[\frac{(1-\Gamma^2)}{2\Gamma^2}
\frac{\sinh\left(z/L \right)}{1-\cosh(1)} \frac{r_{0}(z)}{L} \right]
\label{eq04}
\end{equation}
and
\begin{equation}
B_z(z) \approx B_{z,\mathrm{f}} \left\{ 1+\frac{(1-\Gamma^2)}{\Gamma^2}
\frac{\left[\cosh \left(z/L \right)-\cosh(1)\right]}{1-\cosh (1)}
\right\}. \label{eq05}
\end{equation}
Therefore, $B_r$ and $B_z$ are related by,
\begin{equation}
B_r (z) \approx -\frac{1}{2} r_0 (z) \frac{\mathrm{d} B_z}{\mathrm{d} z}.
\label{eq06}
\end{equation}

\section{Governing equations}
\label{gov}
The ideal MHD equations are linearized by assuming small magnetic perturbations $\mathbi{b}=(b_r, 0, b_z)$ and velocity perturbations
${\boldsymbol \upsilon}=(\upsilon_r, 0, \upsilon_z)$ about a plasma
in static equilibrium. In the derivation we neglect gravity
and assume constant kinetic plasma pressure. Note this means along our model of solar flux tubes that the plasma is not isothermal and the assumption of constant plasma pressure has greater validity in the corona than in the chromosphere. 

Following, e.g., Roberts \& Webb (1978), Roberts (2006), and D\'{i}az \& Roberts (2006), we also neglect the effect of the external environment on the perturbations, i.e., we assume the tube is in a quiescent environment. This means we do not consider any external forces acting on the tube. Inclusion of such effects is essential in the studies of, e.g., \textit{p}-mode absorption of photospheric flux tubes (see Bogdan et al. 1996). The resulting MHD
conservation laws are
\begin{eqnarray}
\rho_{0} \frac{\partial \upsilon_{r}}{\partial t}=-\frac{\partial P_{T}}{\partial %
r}+\frac{1}{\mu} \left( B_{r} \frac{\partial b_{r}}{\partial r}+B_{z}\frac{%
\partial b_{r}}{\partial z} \right) \nonumber \\
+\frac{1}{\mu} \left( b_{r} \frac{\partial B_{r}}{\partial r}+ b_{z}
\frac{\partial B_{r}}{\partial z} \right), \label{eq07} \\
\rho_{0} \frac{\partial \upsilon_{z}}{\partial t}=-\frac{\partial P_{T}}{%
\partial z}+\frac{1}{\mu} \left( B_{r} \frac{\partial b_{z}}{\partial r}%
+B_{z}\frac{\partial b_{z}}{\partial z} \right) \nonumber \\
+\frac{1}{\mu} \left( b_{r} \frac{\partial B_{z}}{\partial r}+b_{z}
\frac{\partial B_{z}}{\partial z} \right), \label{eq08} \\
\frac{\partial b_{r}}{\partial t}=\frac{\partial }{\partial z} \left( B_{z}%
\upsilon_{r}-B_{r} \upsilon_{z} \right), \label{eq09} \\
\frac{\partial b_{z}}{\partial t}=-\frac{1}{r}\frac{\partial }{\partial r} \left[r( %
B_{z} \upsilon_{r}-B_{r} \upsilon_{z}) \right], \label{eq10} \\
\frac{\partial p}{\partial t}=-\gamma p_{0} \frac{1}{r} \frac{\partial %
(r \upsilon_{r})}{\partial r} -\gamma p_{0} \frac{\partial \upsilon_{z}}
{\partial z}, \label{eq11} \\
P_{T}=p+\frac{b_r B_r}{\mu}+\frac{b_z B_z}{\mu}, \label{eq12} \\
\textrm{  and  } \nonumber \\
\frac{1}{r} \frac{\partial }{\partial r} (r b_r) + \frac{\partial b_z}
{\partial z}=0, \label{eq13}
\end{eqnarray}
\noindent where $t$ is time, $r$ and $z$ are the radial and longitudinal
coordinates in the tube, $\rho_0 $ is the plasma mass density in equilibrium,
 $P_T$ is the total perturbation to pressure, $p$ is the kinetic pressure
perturbation, $p_0$ is the kinetic plasma pressure in equilibrium, $B_r$ and $B_z$ are
the background components of the magnetic field, $\gamma$ is the ratio
of specific heats, and $\mu$ is the magnetic permeability.

\subsection{Magnetic Flux Tube Equilibrium}
The potential magnetic field configuration chosen in Section \ref{potential} and our choice of constant plasma pressure put restrictions on the possible types of flux tube equilibria we can model. It is always assumed that the external magnetic field is balanced by the internal one, therefore the flux tube models are not in a magnetic field free environment. Then, our assumption of constant plasma pressure demands that a flux tube with greater internal plasma density than external one must also be cooler than its environment. On the other hand, if the tube is less dense than its environment, the internal temperature must be hotter than the external.

\subsection{Velocity Wave Equation}
Implementing the thin tube approximation we introduce the spatial and temporal scalings $r=\epsilon R$, $z=Z$ and $t=\tau$, where $\epsilon\ll1$. We show the details of our derivation for the governing thin tube $v_z$ velocity equation in Appendix \ref{velocity}. Let us introduce the following Fourier decomposition
$\upsilon_z(Z,\tau)=\upsilon_z(Z) \exp (-i \omega \tau),$
where $\omega$ is the angular frequency of the oscillations. Then, the second-order ordinary differential
equation for the longitudinal velocity wave is
\begin{equation}
\frac{d^2 \upsilon_{z}}{d Z^2}+f_1 (Z) \frac{d \upsilon_z}{d Z} +
\left[ \frac{\omega^2}{c_{\mathrm T}^2}+ f_2 (Z) \right] \upsilon_z = 0,
\label{eq19}
\end{equation}
where
\begin{eqnarray}
f_1 (Z)=\left(\frac{c_{\mathrm s}^{2}-c_{\mathrm A}^{2}}{c_{\mathrm f}%
^{2}} \right) \frac{1}{B_z} \frac{\partial B_z}{\partial Z}, \nonumber
\end{eqnarray}
and
\begin{eqnarray}
f_2 (Z)= -\frac{1}{B_z} \frac{\partial^2 B_{z}}{\partial Z^2}
-\left(\frac{c_{\mathrm s}^{2}-c_{\mathrm A}^{2}}{c_{\mathrm f}^{2}}
\right) \frac{1}{B_z^2} \left( \frac{\partial B_z}{\partial Z}\right)^2 .\nonumber
\end{eqnarray}
For constant magnetic field, Equation (\ref{eq19}) reduces to
\begin{equation}
\frac{d^2 \upsilon_{z}}{d Z^2} + \left(\frac{
\omega^2}{c_{\mathrm T}^2}\right) \upsilon_z =0,
\label{eq20}
\end{equation}
for
\begin{equation}
\frac{\omega^2}{c_{\mathrm T}^2}=\omega^2 \rho_0 \left[ \frac{\mu}{{B_z}^2} +
\frac{1}{\gamma p_0} \right], \label{eq21}
\end{equation}
where the density may depend on $Z$.  This equation (\ref{eq20}) agrees with the
equation obtained by D\'{i}az \& Roberts (2006), where $B_z$ was considered
constant. The solutions to Equation (\ref{eq20}) will depend on the
functional form chosen for $\rho_0$ in Equation (\ref{eq21}).

\subsubsection{Slow Modes in a Homogeneous Tube}
For checking the derivations, we may recover the findings for a homogeneous
tube (e.g., Edwin \& Roberts 1983). In a straight magnetic flux tube
with constant density, the solutions to Equation (\ref{eq20}) are
simply trigonometric functions. Applying the line-tying boundary
at the ends of the tube, $\upsilon_z (\pm L) =0$, we find the
frequencies of the even modes given by
\begin{equation}
\omega_n^e=\frac{c_\mathrm{T}}{L} \frac{(2n-1) \pi}{2},  n=1,2,...
\end{equation}
whereas the odd modes have frequencies given by
\begin{equation}
\omega_n^o=\frac{c_\mathrm{T}}{L} n \pi, n=1,2,...
\end{equation}
The ratio of frequencies of the first overtone and fundamental mode is equal to 2, as expected.

\section{Analytical solution for a smooth density profile}
\label{constantB}
Since we are using cylindrical coordinates we model a solar coronal loop by a straight axisymmetric magnetic flux tube. 
Therefore we neglect the effect of flux tube curvature and are simply modeling the effect of gravitationally 
stratified plasma in a coronal loop which would produce a density profile symmetric about the loop apex. The tube length is 
$2L$ and its radius is $r_0$. The plasma is permeated by a
uniform magnetic field $\mathbi{B}$ directed along the tube axis,
$\mathbi{B}= B_z {\bf \hat{z}}$. The plasma density, $\rho_0 (Z)$, is greater at the loop footpoints than at the apex so this is approximated by the function
   \begin{figure}
   \centering
   \includegraphics[width=8.2cm]{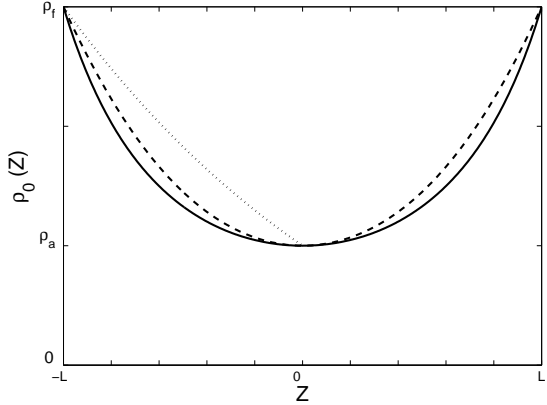}
   \caption{Density profile of the loop. The solid line shows the smooth density profile (\ref{eq33}) while the dashed line shows the
   exponential density profile of a coronal loop (\ref{eq81}), and the dotted line shows the exponential density profile of a chromospheric flux tube (\ref{eq83}).}
              \label{fig2}%
    \end{figure}
\begin{equation}
\rho_0 (Z) = \frac{\rho_{\mathrm a}}{ \left[ 1-(1-\kappa_1)(Z/L)^2
\right]^2}, \label{eq33}
\end{equation}
where $\rho_{\mathrm a} = \rho_{0} (0)$ and
$\rho_{\mathrm f} = \rho_{0} (\pm L)$ are the apex and footpoint
densities, respectively, and
$\kappa_1 = (\rho_{\mathrm a}/\rho_{\mathrm f})^{1/2}$ the stratification
parameter. The solid line in Figure \ref{fig2} shows this smooth density profile as a function of $Z$. In Section 
\ref{numerical}, we choose a more applicable and realistic exponential density profile and solve Equation (\ref{eq20}) 
numerically. However, the choice of density given by Equation (\ref{eq33}) allows us to have a straightforward analytical insight into the effect of density stratification on longitudinal oscillations.

With the equilibrium density profile given by Equation (\ref{eq33}), Equation
(\ref{eq20}) takes the form
\begin{equation}
\frac{d^2 \upsilon_{z}}{d Z^2} +\frac{\omega^2}{c_\mathrm{T,a}^2}
\frac{1}{ \left[ 1-(1-\kappa_1)(Z/L)^2 \right]^2}
\upsilon_z =0, \label{eq34}
\end{equation}
for $c_\mathrm{T,a}= c_\mathrm{T} (0)$, the tube speed at the apex.
The general solution of this equation is (see Polyanin \& Zaitsev 2003)
\begin{equation}
\upsilon_z =  \sqrt{1-(1-\kappa_1)(Z/L)^2} \left( C_1 \cos (u) +C_2
\sin (u) \right), \label{eq35}
\end{equation}
where $C_1$ and $C_2$ are arbitrary constants, and
\begin{equation}
u= \frac{1}{2} \sqrt{\frac{\omega^2 L^2}{c_\mathrm{T,a}^2 (1-\kappa_1)}-1}\left(\ln
\frac{L+\sqrt{1-\kappa_1} Z}{L-\sqrt{1-\kappa_1} Z} \right). \label{eq36}
\end{equation}
To study a standing wave the boundary conditions $\upsilon_z (\pm L) =0$ are
applied, and are satisfied when either $C_2=0$ and $\cos[u(L)]=0$,
or $C_1=0$ and $\sin[u(L)]=0$. The first condition corresponds
to even modes, and the second to odd modes. The frequencies of the even modes are
given by
\begin{eqnarray}
(\omega_n^e)^2= \frac{(1-\kappa_1) c_\mathrm{T,a}^2}{L^2} \times \nonumber \\
 \left[ (2n-1)^2 \pi^2 \left( \ln \frac{1+\sqrt{1-\kappa_1}}{1-\sqrt{1-\kappa_1}}
\right)^{-2} +1 \right], \label{eq37}
\end{eqnarray}
for $n=1,2,...$, while the odd modes have frequencies given by
\begin{eqnarray}
(\omega_n^o)^2= \frac{(1-\kappa_1) c_\mathrm{T,a}^2}{L^2} \times \nonumber \\
\left[ (2 n \pi)^2 \left( \ln \frac{1+\sqrt{1-\kappa_1}}{1-\sqrt{1-\kappa_1}}
\right)^{-2} +1 \right], \label{eq38}
\end{eqnarray}
for $n=1,2,...$

\begin{figure}
   \centering
   \includegraphics[width=7.8cm]{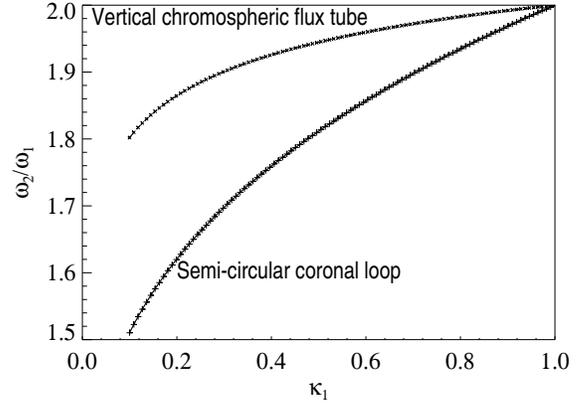}
   \caption{Dependence of the frequency ratio of the first overtone and
   fundamental mode on $\kappa_1$, for a semi-circular coronal loop in solid line (\ref{eq39}),
   and for a vertical chromospheric flux tube in dashed line (\ref{eq78}). The ``$+$'' and ``$x$''
   symbols represent the numerical solutions of Equation (\ref{eq34}) for
   $Z \in [-L,L]$ and $Z \in [-L,0]$, respectively.}
              \label{fig3}%
    \end{figure}

The frequencies of the fundamental mode and the first overtone are given
by Equations (\ref{eq37}) and (\ref{eq38}) with $n=1$, respectively.
Theoretically it is predicted that the frequencies of higher harmonics
have a much stronger dependence on density stratification (e.g., Andries
et al. 2005).
The ratio of frequencies of the first overtone and fundamental mode is given by
\begin{equation}
\frac{\omega_2}{\omega_1}=\frac{\omega_1^o}{\omega_1^e}=\left[\frac{4\pi^2 +\left(\ln\frac
{1+\sqrt{1-\kappa_1}}{1-\sqrt{1-\kappa_1}} \right)^2}{\pi^2 +\left( \ln \frac
{1+\sqrt{1-\kappa_1}}{1-\sqrt{1-\kappa_1}} \right)^2} \right]^{1/2}, \label{eq39}
\end{equation}
which is the same result obtained for the frequency ratio of the transversal mode
by Dymova \& Ruderman (2006).  This is a remarkable property of MHD oscillations
in structured waveguides.  This is also rather assuring, as we arrived to these
results using a completely different approach and modeling.  The dependence of this ratio
of frequencies on $\kappa_1$ is shown in Figure \ref{fig3} by the solid line
(semi-circular coronal loop).  If we consider the limit
of a non-stratified loop, i.e., $\kappa_1 \rightarrow 1$, we find the
frequency ratio tending to 2, and it can be approximated by
\begin{equation}
\frac{\omega_2}{\omega_1}= 2 -\frac{3}{\pi^2}(1-\kappa_1). \label{eq40}
\end{equation}
Equation (\ref{eq40}) shows that the frequency ratio $\omega_2 /\omega_1 <2$
for a stratified loop with constant magnetic field.

\section{Flux tube expansion with constant density}
\label{constantD}
Let us now study the effect of magnetic stratification with a
constant density on the longitudinal oscillations. For this configuration it is more convenient to use the governing pressure perturbation equation. See Appendix \ref{pressure} for insight into the relationship between $v_z (Z)$ and $p(Z)$ when there is longitudinal stratification.
The pressure wave equation (\ref{eq22}) with constant density is
\begin{equation}
\frac{d^2 p}{d Z^2} -\frac{1}{B_z}
\frac{d B_z}{d Z} \frac{d p}{d Z} + \frac{\omega^2}{c_{\mathrm T}^2} p=0. \label{eq41}
\end{equation}
We can transform Equation (\ref{eq41}) to its canonical form using the change
of variable $p = P\sqrt{B_z}$. Then, Equation (\ref{eq41}) becomes
\begin{equation}
\frac{d^2 P}{d Z^2} + \left[ \frac{\omega^2}{c_{\mathrm T}^2}+
\frac{1}{2}\frac{1}{B_z} \frac{d^2 B_z}{d Z^2} -
\frac{3}{4} \frac{1}{B_z^2}
\left(\frac{d B_z}{d Z}\right)^2\right] P=0. \label{eq42}
\end{equation}
For weak magnetic stratification the $\omega^2 / c_{\mathrm T}^2$ term has a dominant
effect on the eigenvalues. To analytically investigate the behavior of the
eigenvalues in this regime, we approximate Equation (\ref{eq42}) with
\begin{equation}
\frac{d^2 P}{d Z^2} +\frac{\omega^2}{c_{\mathrm T}^2} P=0. \label{eq43}
\end{equation}
We suggest to use a rational function for the tube speed $c_{\mathrm T}$ defined by
\begin{equation}
c_{\mathrm T}^2 (Z) = c_\mathrm{T,a}^2 \left[ 1+
\left(\frac{c_\mathrm{T,f}}{c_\mathrm{T,a}} -1\right)
\left(\frac{Z}{L}\right)^2 \right]^2, \label{eq67}
\end{equation}
for $c_\mathrm{T,a} = c_\mathrm{T} (0)$ and
$c_\mathrm{T,f}= c_\mathrm{T} (\pm L)$ being
the apex and footpoint tube speeds, respectively.
In Section \ref{numerical}, we solve the governing velocity equation (\ref{eq19}) with our potential 
field definition of $B_z(z)$ from Equation (\ref{eq05}) numerically. However, the choice of tube 
speed given by Equation (\ref{eq67}) allows us to have a straightforward analytical insight into the effect of magnetic stratification on longitudinal oscillations.

An exact solution to Equation (\ref{eq43}) with $c_{\mathrm T}$ defined
by Equation (\ref{eq67}) is given by (see Polyanin \& Zaitsev 2003)
\begin{equation}
P =  \sqrt{ \alpha^2 Z^2 +1} \left( C_3 \cos (\nu) +C_4 \sin (\nu)
\right), \label{eq69}
\end{equation}
where $C_3$ and $C_4$ are arbitrary constants, with
\begin{equation}
\nu= \sqrt{ \frac{\omega^2}{\alpha^2 c_\mathrm{T,a}^2}+1} \arctan
(\alpha Z), \label{eq70}
\end{equation}
and
\begin{equation}
\alpha = \frac{1}{L} \sqrt{\frac{c_\mathrm{T,f}}{c_\mathrm{T,a}} -1}.
\end{equation}
In this case, to find the standing mode solution the same boundary conditions
$\upsilon_z (\pm L)= P' (\pm L) =0$ are applied. The frequencies of the even modes
are given by
\begin{equation}
\frac{(\omega_n^e) L}{c_\mathrm{T,a}} = n \pi +\left(\frac{n\pi}{3} +\frac{1}{2n\pi}\right)(\alpha L)^{2} +O (\alpha L)^{4}, \label{eq72}
\end{equation}
for $n=1,2,...$, while the frequencies of the odd modes are
\begin{eqnarray}
\frac{(\omega_n^o) L}{c_\mathrm{T,a}} & = & (n-\frac{1}{2}) \pi +\left(\frac{ (n-\frac{1}{2})\pi}{3} +\frac{1}{2(n-\frac{1}{2})\pi}\right)(\alpha L)^{2}
 \nonumber \\
& &  +O (\alpha L)^{4}, 
\label{eq71}
\end{eqnarray}
for $n=1,2,...$.  The ratio of frequencies of the first overtone and fundamental mode is
   \begin{figure}
   \centering
   \includegraphics[width=7.8cm]{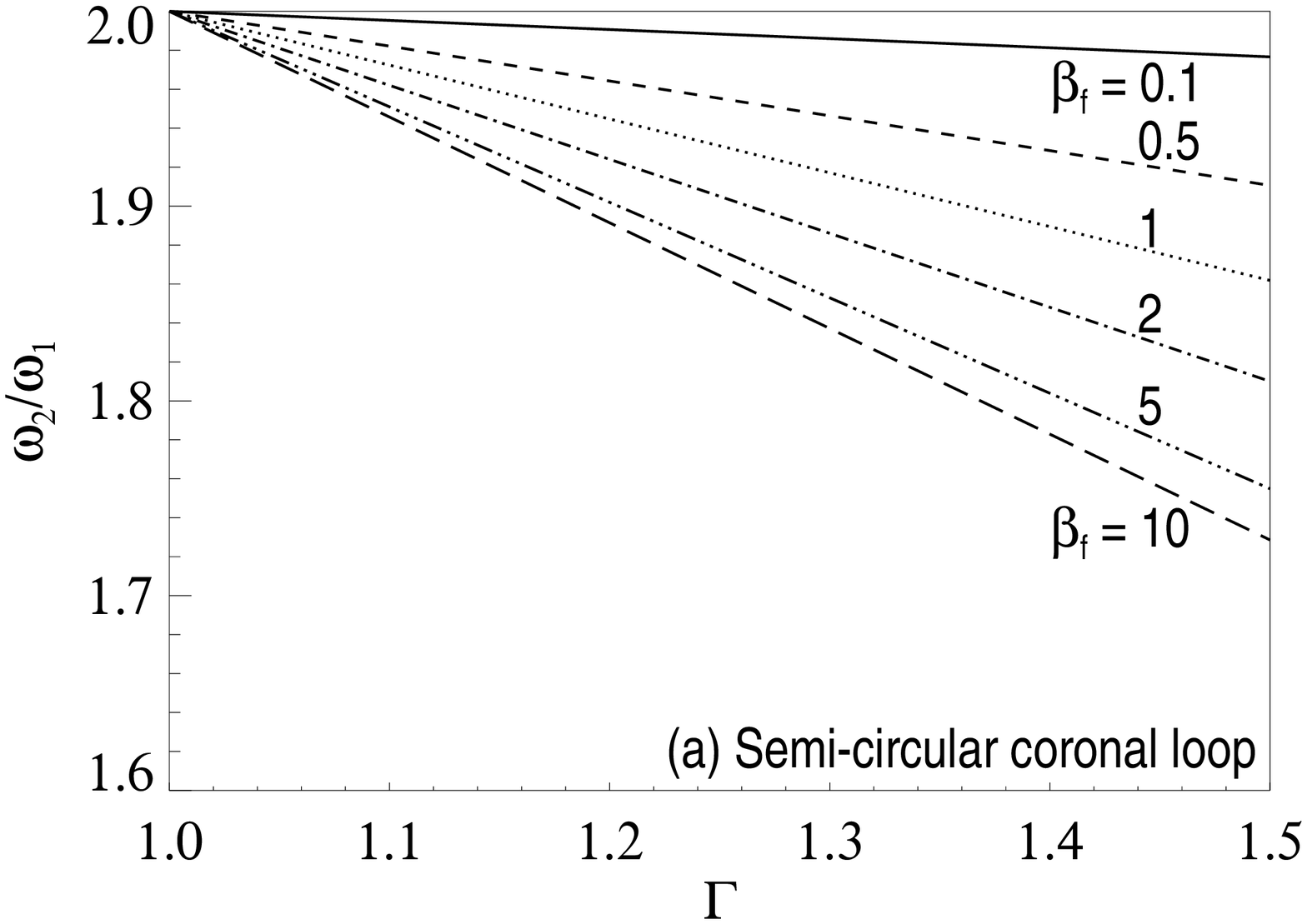}
   \includegraphics[width=7.8cm]{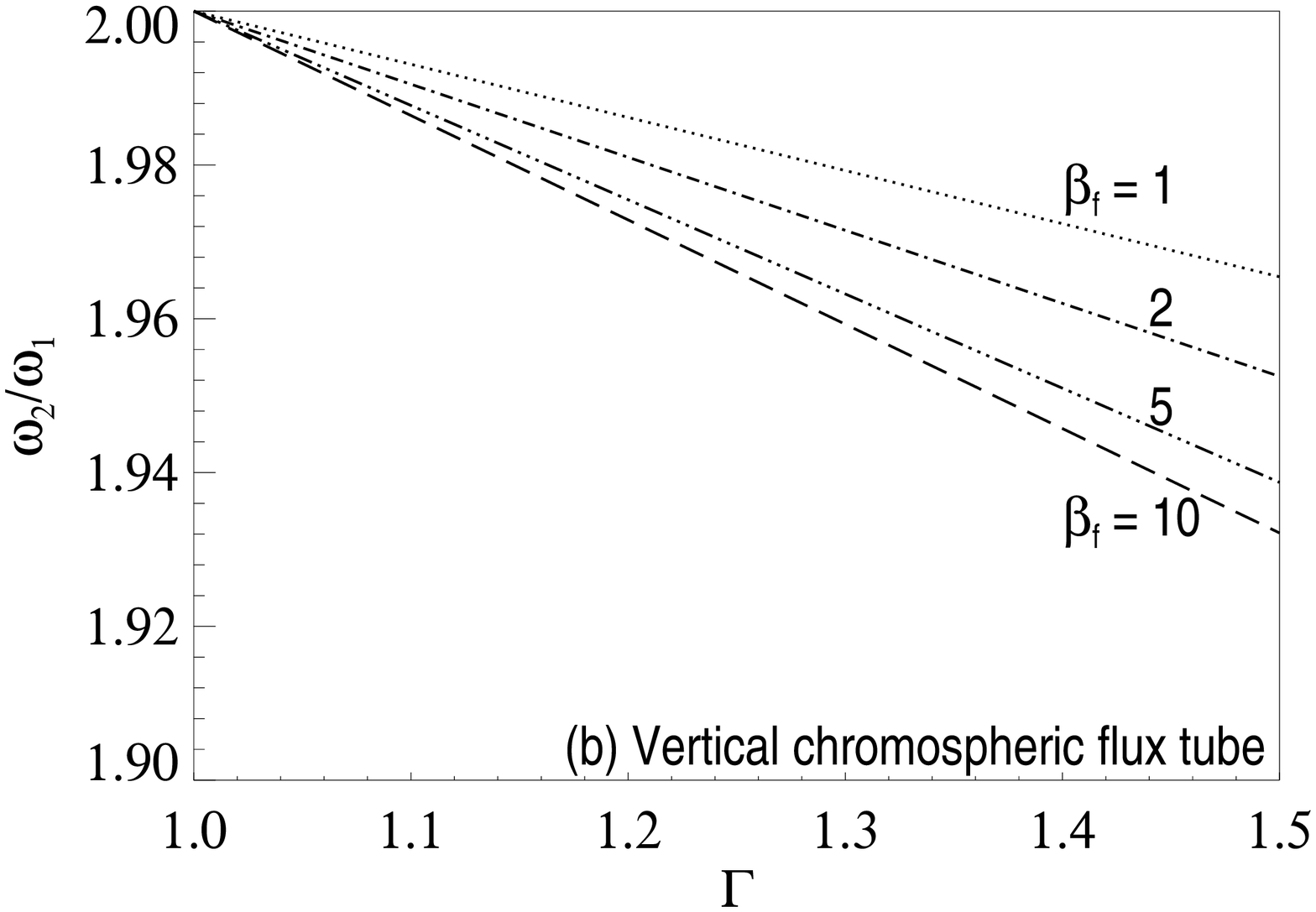}
   \caption{Dependence of the ratio of frequencies of the first overtone and
   fundamental mode (Equations (\ref{eq75}) and (\ref{eq80})) on the expansion parameter $\Gamma$ for
   different values of $\beta_{\mathrm f}$. Solid, dashed, dotted, dot-dashed, triple-dot-dashed,
   and long-dashed lines correspond to $\beta_{\mathrm f} =0.1$, 0.5, 1,
   2, 5, and 10, respectively.}
              \label{fig4}%
    \end{figure}

\begin{equation}
\frac{\omega_2}{\omega_1}=\frac{\omega_1^e}{\omega_1^o} =2 -\frac{3}{\pi^2} (\alpha L)^2+O (\alpha L)^4. \label{eq73}
\end{equation}
We now want to see how the effect of $\Gamma$ is related to plasma $\beta$. Since we have not defined $\Gamma$ explicitly 
in Equation (\ref{eq67}), we combine this definition of tube speed with our potential magnetic field of $B_z(z)$ given by Equation 
(\ref{eq05}), which results in the ratio of tube speeds,
\begin{equation}
\frac{c_\mathrm{T,f}}{c_\mathrm{T,a}} = \left[ \frac
{2+ \gamma \beta_{\mathrm f} \Gamma^4 }
{2+ \gamma \beta_{\mathrm f}}  \right]^{1/2}, \label{eq74}
\end{equation}
where $\beta_{\mathrm f}= 2 \mu p_0 / B_\mathrm{z,f}^2$ is the beta plasma
at the footpoints. Therefore, Equation (\ref{eq73}) shows that change in
frequency ratio is dependent only on the loop expansion factor $\Gamma$ for
any finite beta plasma.

If we consider the limit of a non-expanding loop, i.e., $\Gamma \sim1$,
we recover a constant tube speed along the loop.
The ratio of frequencies can be Taylor expanded for $\Gamma \sim 1$ and $\gamma=5/3$, giving
\begin{equation}
\frac{\omega_2}{\omega_1}= 2 -\frac{30 \beta_{\mathrm f}}{(6+5\beta_{\mathrm
f})\pi^2} (\Gamma-1). \label{eq75}
\end{equation}
Equation (\ref{eq75}) clearly shows that $\omega_2 /\omega_1 < 2$ for
oscillations in an expanding magnetic flux tube with constant density, and
it shows a clearly linear dependence with the expansion factor. This relationship between
$\omega_2 /\omega_1$ and $\Gamma$ is shown in Figure \ref{fig4}(a).
It is clear that increasing the magnetic stratification takes to a lower frequency ratio,
and this effect is more significant for plasmas with higher
$\beta_{\mathrm f}$.

\section{Application to Solar Physics}
Our results are relevant to magneto-seismology, e.g., estimating the
coronal density scale height by using the observed ratio of the fundamental
frequency and first overtone of longitudinal loop oscillations.

\subsection{Application to the Corona}
In the solar corona the thermal pressure is generally smaller than the magnetic
pressure, giving a plasma $\beta$ parameter $\ll 1$.
Stratified coronal loops ($0< \kappa_1 <1$) with constant magnetic field
give a frequency ratio lower than 2 (Equation (\ref{eq40})).

For a 1 MK average corona, the expected hydrostatic scale height $H$ should
be about 50 Mm, therefore for any coronal loop with length of the order of
$H$ or less could be approximated by the configuration in Section 5, where
the density is constant inside the loop and the tube speed (i.e., geometry
of the magnetic field) is given by Equation (\ref{eq67}).

For a very small beta plasma (i.e., $0.01<\beta<0.1$)
the expansion has a weak effect on the frequency ratio, as it can be
seen in Figure \ref{fig4}(a) for the solid line. The frequency ratio can be Taylor
expanded for a very small plasma $\beta_{\mathrm f}$ parameter
($\beta_{\mathrm f} \ll 1$), giving
\begin{equation}
\frac{\omega_2}{\omega_1}= 2- \frac{5}{4} \left( \frac{\Gamma^4-1}{\pi^2}\right)
\beta_{\mathrm f}.  \label{eq76}
\end{equation}
Equation (\ref{eq76}) shows that $\omega_2 /\omega_1 <2$ for an expanding
magnetic flux tube ($\Gamma>1$) in the corona with constant density.

\subsection{Lower Solar Atmosphere}
We now calculate the frequencies of standing modes in an expanding chromospheric
($\beta >1$) flux tube with fixed boundaries at the photosphere and transition
region, since, e.g., Fujimura \& Tsuneta (2009) have recently obtained observational
evidence of such waves. Note that it could be possible that flux tubes undergo large expansions in the chromosphere (e.g., Tsuneta et al. 2008), so our assumption of weak magnetic stratification may have limitations in application to chromospheric wave observations. However, the size of corrections to eigenfrequencies if larger flux tube expansions are considered are as of yet still unquantified and must be the focus of a future study. Regarding our current model, to calculate the eigenfrequencies of a vertical chromospheric weakly expanding flux tube,
we need to solve the eigenvalue problem in only half of our magnetic bottle, and
therefore, the boundary conditions $\upsilon_z (-L) =0$ and $\upsilon_z (0) =0$
are applied.

For the loop with density stratification and constant magnetic field, it is
found that $C_1=0$ and $\sin[u(-L)]=0$ for any arbitrary constant $C_2$.
The frequencies for all odd and even modes ($\omega_n$) are given by Equation
(\ref{eq38}).
The frequencies of the fundamental mode and the first overtone are given by
Equation (\ref{eq38}) with $n=1$ and 2, respectively. The ratio of frequencies
of the first overtone and fundamental mode is
\begin{equation}
\frac{\omega_2}{\omega_1}= \left[ \frac{16 \pi^2 + \left( \ln\frac
{1-\sqrt{1-\kappa_1}}{1+\sqrt{1-\kappa_1}} \right)^2}{4 \pi^2 +\left( \ln \frac
{1-\sqrt{1-\kappa_1}}{1+\sqrt{1-\kappa_1}} \right)^2} \right]^{1/2} . \label{eq78}
\end{equation}
It is important to note, that the value of $\omega_2/\omega_1$ strongly
depends on which functional form is chosen for the equilibrium density. Figure \ref{fig3} shows
a comparison of the dependence of the frequency ratios on the parameter $\kappa_1$,
in the lower atmosphere (dashed line) and in the corona (solid line). When we
consider the limit of a non-stratified loop, i.e., $\kappa_1=1$,
we recover the ratio of
frequencies equal to 2. Equation (\ref{eq78}) can be approximated to
\begin{equation}
\frac{\omega_2}{\omega_1}= 2 -\frac{3}{4 \pi^2}(1-\kappa_1), \label{eq78a}
\end{equation}
for $\kappa_1 \sim 1$.
Equation (\ref{eq78a}) shows that the frequency ratio $\omega_2 /\omega_1 <2$
for a stratified loop ($\kappa_1 <1$) in the lower solar atmosphere. Note that for our coronal loop model the densities at both tube ends are equal with the density profile symmetric about the apex. In our vertical chromospheric flux tube model the densities at both ends are not equal, and the density is monotonically decreasing as a function of height. It is found that the effect of the density stratification on the frequency
ratio is larger in the corona than in the lower atmosphere, due to the
asymmetric nature of the density profile in the chromosphere.

In the second case, a loop with magnetic stratification and constant density, the
same boundary conditions are applied.
The frequencies for all odd and even modes $(\omega_n)$ are given by Equation
(\ref{eq72}).

The ratio of frequencies of the first overtone and fundamental mode is
\begin{equation}
\frac{\omega_2}{\omega_1}= 2-\frac{3}{4\pi^2} (\alpha L)^2+O (\alpha L)^4. \label{eq79}
\end{equation}
The Taylor expansion of Equation (\ref{eq79}) for $\Gamma \sim 1$ is
\begin{equation}
\frac{\omega_2}{\omega_1}= 2 - \frac{15}{2} \frac{\beta_{\mathrm f}}{(6+5
\beta_{\mathrm f}) \pi^2} (\Gamma-1), \label{eq80}
\end{equation}
showing again that $\omega_2 /\omega_1 <2$ and it has a linear dependence
on $\Gamma$ for standing waves in the lower atmosphere with
constant density and weak magnetic expansion.
This frequency ratio as a function of $\Gamma$ is shown in Figure \ref{fig4}(b).
The effect of the magnetic stratification on the frequency ratio is smaller
in the lower solar atmosphere than in the corona.

\section{Numerical solutions}
\label{numerical}
In this section, we compare the analytical approximate solutions with
the numerical solution of Equation (\ref{eq19}), using the shooting method
based on the Runge-Kutta technique.

\subsection{Stratified Loop with Constant Magnetic Field}
Equation (\ref{eq19}) becomes Equation (\ref{eq20}) when
$B_z$ is constant.
Equation (\ref{eq20}) can be solved numerically for both a semi-circular
coronal loop and a vertical chromospheric flux tube, depending on the equilibrium chosen.
   \begin{figure}
   \centering
   \includegraphics[width=7.8cm]{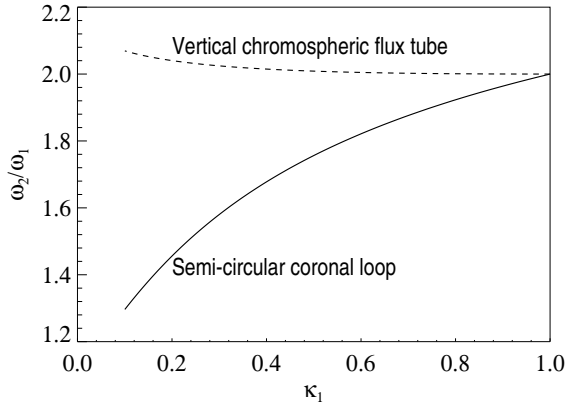}
   \caption{Dependence of the ratio of frequencies of the first overtone and
   fundamental mode on $\kappa_1$, for a semi-circular coronal loop in solid line
   (\ref{eq81}) and for a vertical chromospheric flux tube in dashed line (\ref{eq83}),
   obtained from the numerical calculations.}
              \label{fig7}%
    \end{figure}

In the first case, for a semi-circular coronal loop perpendicular to the
plane of the photosphere with $z \in [-L,L]$, the density is defined by
(see dashed line of Figure \ref{fig2})
\begin{equation}
\rho_0 (z) = \rho_\textrm{f} \exp \left[ -\frac{2L}{\pi H} \cos
\left(\frac{\pi z}{2L}\right)\right], \label{eq81}
\end{equation}
\noindent where $H$ is the density scale height and $\rho_\textrm{f}$
the footpoint density. Hence, in this case the parameter $\kappa_1$ of Section
4 is a function of $H$ and is given by
\begin{equation}
\kappa_1^2 =\exp \left( -\frac{2L}{\pi H}\right). \label{eq82}
\end{equation}
The frequency ratio obtained with the density profile (\ref{eq81}) as a
function of $\kappa_1$ is shown in Figure \ref{fig7} by the solid line. It is
clearly lower than 2, similar to the results obtained in the analytical
case.

For a vertical chromospheric flux tube with $z\in [-L, 0]$, the density
profile shown by the dotted line in Figure \ref{fig2} is
\begin{equation}
\rho_0 (z) = \rho_\textrm{f} \exp \left[ -\frac{(z+L)}{H} \right], \label{eq83}
\end{equation}
where the parameter $\kappa_1$ is now given by
\begin{equation}
\kappa_1^2 =\exp \left( -\frac{L}{H}\right). \label{eq84}
\end{equation}
Equation (\ref{eq20}) with density profile given by Equation (\ref{eq83}) has the well-known solution
\begin{equation}
v_z(Z)=C_5 \mathrm{J_0}\left(\frac{2H \omega}{c_T(Z)} \right)+C_6\mathrm{Y_0}\left(\frac{2 H \omega}{c_T(Z)} \right),
\label{bessel}
\end{equation}
where $C_5$ and $C_6$ are arbitrary constants. To find the eigenvalues using solution (\ref{bessel}), 
we must solve a transcendental equation by either analytical or numerical techniques. Here, we choose to solve 
it numerically. See, e.g., McEwan et al. (2008) and Verth et al. (2010) for analytical solutions of equivalent equations in their studies of other MHD waves in limiting cases of weak stratification for both coronal loop and vertical flux tube geometries.

The dependence of the ratio of frequencies on $\kappa_1$ for the density
profile (\ref{eq83}) is shown in Figure \ref{fig7} by the dashed line.
In this case, the frequency ratio is greater than 2, confirming that this
parameter strongly depends on the choice of the functional form of density.
This means that caution must be used when interpreting the frequency ratio of chromospheric 
standing modes. For example, the choice of a density profile that gives a tube speed increasing linearly with height results in $\omega_2/\omega_1<2$, while the choice of a density profile giving a tube speed exponentially increasing with height gives $\omega_2/\omega_1>2$.

\subsection{Expanding Loop with Constant Density}

\begin{figure}
   \centering
   \includegraphics[width=7.8cm]{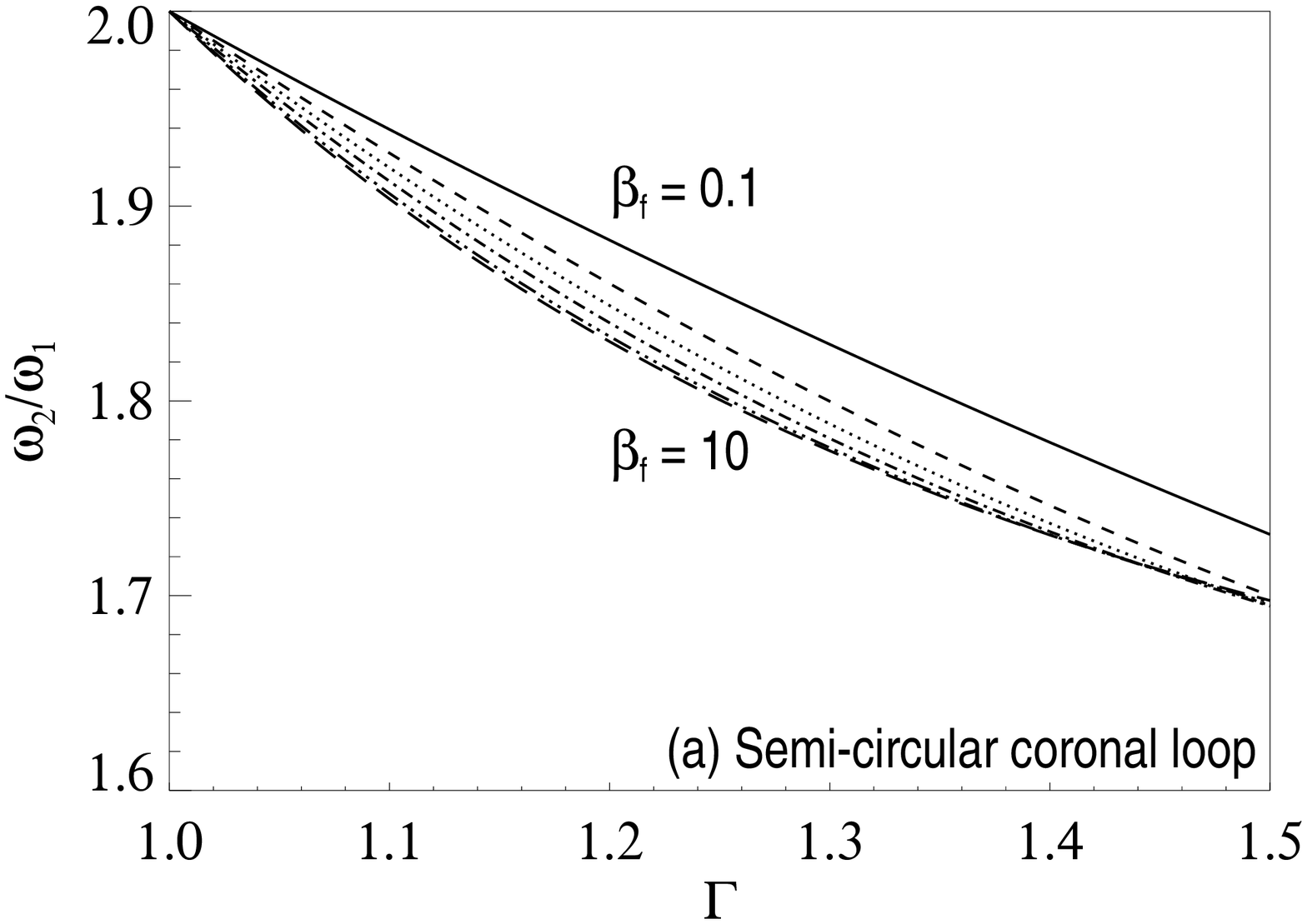}
   \includegraphics[width=7.8cm]{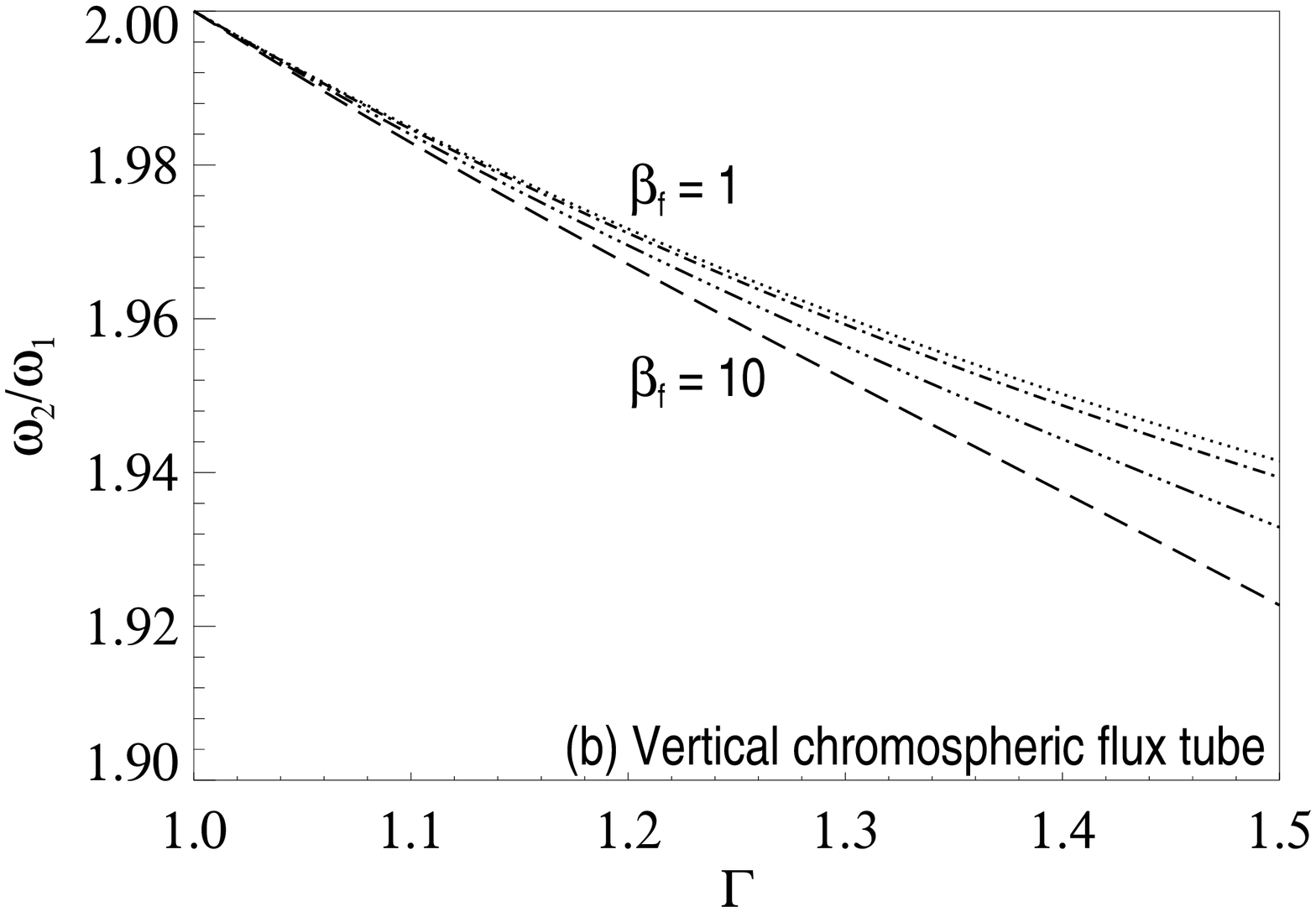}
   \caption{Dependence of the frequency ratio of the first overtone and
   fundamental mode on the expansion parameter $\Gamma$ for
   different values of $\beta_{\mathrm f}$ obtained from the numerical
   calculations. Solid, dashed, dotted, dot-dashed, triple-dot-dashed,
   and long-dashed lines correspond to $\beta_{\mathrm f} =0.1$, 0.5, 1,
   2, 5, and 10, respectively.}
              \label{fig8}%
    \end{figure}

Using Equation (\ref{eq05}) for $B_z(z)$ with $z \in [-L,L]$ we can now compute the
numerical solution of Equation (\ref{eq19}) for longitudinal oscillations in
a coronal loop.  Figure \ref{fig8}(a) shows the frequency ratio of the first
overtone and fundamental mode as a function of the expansion parameter $\Gamma$
for different values of $\beta_{\mathrm f}$.  It is found that increasing the
magnetic stratification leads to a lower frequency ratio,
and this effect is more significant for solar waveguides with
higher $\beta_{\mathrm f}$.
These results are very similar to the results obtained in the analytical analysis.


To study the eigenmodes of a vertical chromospheric flux
tube, we use Equation (\ref{eq05}) for $B_z(z)$ with $z\in [-L,0]$ obtaining the
numerical solution of Equation (\ref{eq19}). The dependence of the
frequency ratio of the first overtone and fundamental mode on the
expansion parameter is shown in
Figure \ref{fig8}(b).  The effects of the expansion $\Gamma$ is somewhat
smaller in
the chromosphere than in the corona, being consistent with the analytical results.
Also, the values of the frequency ratio for $\Gamma <1.5$ (weak expansion)
obtained from numeric and analytical calculations are very similar.

\section{Implications for solar magneto-seismology}
The governing equation (\ref{eq19}) is valid for both density and magnetic stratification. We investigated the two effects, density and 
magnetic variations, on the value of $\omega_2/\omega_1$ separately for clarity in Sections \ref{constantB} - 7. 
Now we discuss the combined effects in realistic solar waveguides where it is likely that the two types of stratification are present 
simultaneously.

\subsection{Semi-circular Coronal Loops: Effect of both Density and Magnetic Stratification}
In the case of a coronal loop, it was shown in Sections \ref{constantB} and \ref{constantD} analytically and Section \ref{numerical} 
numerically, that if there is density stratification (magnetic field constant) then $\omega_2/\omega_1<2$ and if there is magnetic 
stratification (constant density) then also $\omega_2/\omega_1<2$. Hence, regarding the value of $\omega_2/\omega_1$ the two 
effects are not competing against each other, in contrast to the kink mode (see, e.g., Verth \& Erd\'{e}lyi 2008; Ruderman et al. 2008). 
This result is robust, in particular, in the functional forms chosen for $\rho_0(Z)$ and $B_z(Z)$ as long as there is symmetry about 
the loop apex.

Observational claims of estimating slow mode values of $\omega_2/\omega_1$ from intensity perturbations in coronal loops have 
been made using both EUV (Srivastava \& Dwivedi 2010) and X-ray (Kumar et al. 2011) data from {\it Hinode}. Srivastava \& Dwivedi (2010) 
claim to detect two separate cases in EUV coronal loops of $\omega_2/\omega_1<2$. However, in these observed coronal loops the 
value of plasma $\beta$ may be very low, and in this limit $c_T(z) \approx c_s(Z)$. In this situation it is the variation of $c_s(Z)$ in wave 
equation (\ref{eq19}) that would have the dominating effect on the value of $\omega_2/\omega_1$ and magnetic stratification, even if 
present, would not play a significant part. Assuming the observed loops are isobaric and density is decreasing as a function of height, 
then even in the zero-$\beta$ limit, our model gives $\omega_2/\omega_1<2$, consistent with these particular EUV observations. 
Physically, this would be due to $c_s(Z)$ increasing with height in a coronal loop, i.e., temperature hotter at the apex than at the footpoints.

Regarding the study by  Kumar et al. (2011) of slow mode values of $\omega_2/\omega_1$ in X-ray small-scale loops, Table 1 of that 
paper shows six separate estimates of $\omega_2/\omega_1$ and one particular case of $\omega_3/\omega_1$. Three estimates have 
$\omega_2/\omega_1<2$, similar to the measurements of Srivastava \& Dwivedi (2010), but in contrast three also have 
$\omega_2/\omega_1>2$. In the low-$\beta$ limit, the observations of $\omega_2/\omega_1>2$, could be explained by $c_s(Z)$ 
decreasing as a function of height in these loops, i.e., the temperature being hotter at foot-points than at the apex. However, to explain 
this our isobaric model would require density increasing with height to have $c_s(Z)$ decreasing with height and this is unphysical. 
To model such a coronal loop more realistically, one would also need to model $p(Z)$ and choose $\rho(Z)$ such that temperature 
is hotter at foot-points than at apex. This would be an important future extension to this current work.

\subsection{Vertical Flux Tubes: Effect of both Density and Magnetic Stratification}

Our study of magnetic stratification (constant density) in vertical flux tubes resulted in $\omega_2/\omega_1<2$ for both the hyperbolic 
profile for $B_z(Z)$ given by Equation (\ref{eq05}) and the tube speed given by the rational function in Equation (\ref{eq67}). However, 
it was shown that for the case of density stratification (constant magnetic field) in a vertical flux tube, $\omega_2/\omega_1$ could 
either be greater or less than 2 depending on the functional form chosen.  E.g., $\omega_2/\omega_1<2$ for the smooth rational 
function of density given by Equation (\ref{eq33}) and $\omega_2/\omega_1>2$ for the exponential profile of $\rho_0(Z)$ given by 
Equation (\ref{eq83}). In our investigation we have shown that, depending on the functional forms chosen, density stratification and 
magnetic stratification could both cause the value of $\omega_2/\omega_1$ to be less than 2, or the effects could oppose each other. 
This means that magneto-seismology for vertical flux tubes in the chromosphere may be more subtle than in the case of coronal loops 
when it comes to interpreting the observed values of $\omega_2/\omega_1$.

\section{Summary and conclusions}

In this paper the effects of density and magnetic stratification on
longitudinal oscillations in isobaric coronal and chromospheric conditions have been studied.
Solar waveguides were modeled as axisymmetric cylindrical magnetic tubes. The governing equation were derived and solved
by analytical approximation, and examined in the thin flux tube
limit. We studied the effects
of magnetic stratification while the density is constant and density
stratification for a constant magnetic field.

From the analytic solutions, both density stratified
and expanding coronal loops have $\omega_2 /\omega_1 <2$. For small expansion, a linear dependence between
the frequency ratio and the expansion factor is found.
It was also found that the effect of magnetic field strength decreasing with
height has the same effect on the frequency ratios to that of
gravitational density stratification, in contrast with the results
for kink modes (Verth \& Erd\'elyi 2008; Andries et al. 2005).

It was found that the introduction of waveguide structuring
results in a modification to the oscillatory frequency of the
mode. The expression for the frequency ratio obtained in Section 5 (Equation (\ref{eq75}))
depends on the expansion parameter $\Gamma$ and an additional dependence on $\beta_{\mathrm f}$ was found, due to the inclusion of kinetic pressure in our model.

Next, numerical solutions were performed. The numerical
results were consistent with the analytic solutions for the coronal loop model.
Also, for the vertical chromospheric flux tube model it was found that the frequency ratio strongly depends on the
functional form of the density, suggesting that caution must be used when interpreting the frequency ratios of chromospheric standing modes.

The effect of gravity was neglected in the present study. It is known for propagating longitudinal waves that the cutoff 
frequency is increased by the inclusion of gravity (see, e.g., Roberts \& Webb 1978), which introduces a 
Brunt-V\"{a}is\"{a}l\"{a} (buoyancy) term into the governing equations. This term has been shown to be relevant for the 
leakage of \textit{p}-mode driven longitudinal waves into the upper atmosphere (De Pontieu et al. 2004, 2005), since the 
influence of the Brunt-V\"{a}is\"{a}l\"{a} term in the cutoff frequency can be reduced by the amount of magnetic field inclination, i.e., the more tilted the field from the vertical, the lower the effective cut-off frequency. In a future study, it would therefore be of great interest to investigate the effect of gravity and field inclination on the longitudinal standing modes, e.g., to quantify the importance of the Brunt-V\"{a}is\"{a}l\"{a} (buoyancy) term on the eigenfunctions and eigenvalues.

It is important to progress the field of magneto-seismology that any proposed model can be
tested against the observed oscillatory properties of solar waveguides. The theory presented in this paper should be 
helpful in this regard, modeling both magnetic and density stratification along solar waveguides which can be used to interpret observations of standing slow oscillations such as those by Srivastava \& Dwivedi (2010) and Kumar et al. (2011). In the future, it will also be beneficial to find the solutions of the governing equation when stronger magnetic stratification is considered.

\acknowledgments
M.L.-C. thanks the support by the University of Sheffield Fellowship,
Overseas Research Students Award (ORS), and the financial support from the Argentinean grant PICT 2007-1790 (ANPCyT).
R.E. acknowledges M. K\'eray for patient encouragement and is also
grateful to NSF, Hungary (OTKA, Ref. No. K83133).
The authors also thank R. Morton for providing guidance on the numerical results.


\appendix
\section{Derivation of velocity wave equation}
\label{velocity}
In this Appendix we derive the velocity wave equation (\ref{eq19}).
Equations (\ref{eq07})-(\ref{eq13}) may be combined to yield the
general equations for $\upsilon_z$, $\upsilon_r$, and $P_T$, 
\begin{figure*}[h]
\begin{eqnarray}
\frac{\partial }{\partial z} \left(\frac{\partial P_T}{\partial t}\right) +
\left[ \rho_{0} \frac{\partial^2 \upsilon_{z}}{\partial t^2}-
\frac{B_r^2}{\mu} \frac{\partial^2 \upsilon_{z}}{\partial r^2}-
\frac{B_r B_z}{\mu} \frac{\partial^2 \upsilon_{z}}{\partial r \partial z} \right] 
+\left[-\frac{B_r}{\mu} \frac{\partial B_r}{\partial r}
-\frac{B_z}{\mu} \frac{\partial B_r}{\partial z}\right]\frac{\partial \upsilon_z}{\partial r} 
+\left[ \frac{B_z}{\mu} \frac{\partial B_z}{\partial z} +\frac{B_r}{\mu} \frac{\partial B_z}
{\partial r}\right] \frac{\partial \upsilon_z}{\partial z} \nonumber \\
+\left[ \frac{B_r}{\mu} \frac{\partial^2 B_{z}}{\partial r \partial z}+
\frac{1}{\mu} \frac{\partial B_z}{\partial r}\frac{\partial B_r}{\partial z}
+\frac{B_z}{\mu} \frac{\partial^2 B_{z}}{\partial z^2} +\frac{1}{\mu} \left(\frac{\partial B_z}
{\partial z}\right)^2 \right] \upsilon_z 
+\left[ \frac{B_r B_z}{\mu} \frac{\partial^2 \upsilon_{r}}{\partial r^2} +
\frac{B_z^2}{\mu} \frac{\partial^2 \upsilon_{r}}{\partial r \partial z} \right]
+\frac{B_z^2}{\mu r} \frac{\partial \upsilon_r}{\partial z} \nonumber \\
+\left[ \frac{2 B_r}{\mu} \frac{\partial B_z}{\partial r} +\frac{B_r B_z}{\mu r} +
\frac{2 B_z}{\mu} \frac{\partial B_z}{\partial z} \right]
\frac{\partial \upsilon_r}{\partial r} 
+\left[ \frac{B_r}{\mu r} \frac{\partial B_z}{\partial r} +
\frac{B_r}{\mu} \frac{\partial^2 B_z}{\partial r^2} +
\frac{B_z}{\mu} \frac{\partial^2 B_z}{\partial r \partial z} - \frac{B_r B_z}{\mu r^2}+
\frac{2 B_z}{\mu r} \frac{\partial B_z}{\partial z} \right] \upsilon_r=0, \label{eq14} \\
\frac{\partial }{\partial r} \left(\frac{\partial P_T}{\partial t}\right) +
\left[ \rho_{0} \frac{\partial^2 \upsilon_{r}}{\partial t^2}- \frac{B_z^2}{\mu}
\frac{\partial^2 \upsilon_{r}}{\partial z^2} - \frac{B_r B_z}{\mu}
\frac{\partial^2 \upsilon_{r}}{\partial r \partial z} \right] 
+\left[ \frac{B_z}{\mu} \frac{\partial B_r}{\partial z} -\frac{B_r}{\mu}
\frac{\partial B_z}{\partial z} \right] \frac{\partial \upsilon_r}{\partial r} \nonumber \\
-\left[ \frac{B_r}{\mu} \frac{\partial B_z}{\partial r}+
\frac{2 B_z}{\mu} \frac{\partial B_z}{\partial z} +\frac{B_z}{\mu}
\frac{\partial B_r}{\partial r} \right] \frac{\partial \upsilon_r}{\partial z} 
+\left[ \frac{1}{\mu} \frac{\partial B_z}{\partial r} \frac{\partial B_r}{\partial z}
-\frac{B_r}{\mu}\frac{\partial^2 B_z}{\partial r \partial z}
-\frac{B_z}{\mu}\frac{\partial^2 B_z}{\partial z^2}
+\frac{B_z}{\mu r} \frac{\partial B_r}{\partial z}
-\frac{1}{\mu} \frac{\partial B_r}{\partial r} \frac{\partial B_z}{\partial z}
\right] \upsilon_r \nonumber \\
+\left[ \frac{B_r^2}{\mu} \frac{\partial^2 \upsilon_{z}}{\partial r \partial z}
+\frac{B_r B_z}{\mu} \frac{\partial^2 \upsilon_{z}}{\partial z^2} \right]
+\left[ \frac{2 B_r}{\mu} \frac{\partial B_r}{\partial r} +\frac{2 B_z}{\mu}
\frac{\partial B_r}{\partial z} \right] \frac{\partial \upsilon_z}{\partial z} 
+\left[ \frac{B_r}{\mu} \frac{\partial^2 B_r}{\partial r \partial z}
+\frac{B_z}{\mu} \frac{\partial^2 B_r}{\partial z^2} -
\frac{B_r}{\mu r} \frac{\partial B_r}{\partial z} \right] \upsilon_z = 0, \label{eq15} \\
\frac{\partial P_T}{\partial t}= -\frac{B_z^2}{\mu} \left( \frac{\partial
\upsilon_r}{\partial r} +\frac{\upsilon_r}{r} \right)- \gamma p_0
\left( \nabla \cdot {\boldsymbol \upsilon} \right) 
+\left(\frac{B_r B_z}{\mu}\frac{\partial \upsilon_z}{\partial r}
-\frac{B_r^2}{\mu}\frac{\partial \upsilon_z}{\partial z} \right) \nonumber \\
-\left(\frac{B_r}{\mu} \frac{\partial B_r}{\partial z}
+\frac{B_z}{\mu} \frac{\partial B_z}{\partial z}\right) \upsilon_z  
+\frac{B_r B_z}{\mu} \frac{\partial \upsilon_r}{\partial z}
+\left(\frac{B_r}{\mu} \frac{\partial B_z}{\partial z}
-\frac{B_z}{\mu} \frac{\partial B_z}{\partial r}\right)\upsilon_r,
\label{eq16}
\end{eqnarray}
\end{figure*}

If we compare Equations (\ref{eq14})-(\ref{eq16}) with Equations (1)-(3) from D\'{i}az \& Roberts (2006, who studied 
density stratification with a constant magnetic field), the added complexity to the governing wave equations if we have an equilibrium with an expanding magnetic field can readily be appreciated.

Here we are interested in the behavior of slow magnetoacoustic modes, we introduce
the following spatial and temporal scalings:
\begin{eqnarray}
r=\epsilon R,   \, \; z=Z,   \,  \; \upsilon_r=\epsilon \upsilon_R,   \,  \;
\upsilon_z=\upsilon_z, 
B_r=\epsilon B_R,  \,  \; B_z=B_z,  \,  \; \textrm{and } \,  \; t=\tau. \label{eq17}
\end{eqnarray}
It is interesting to note that such scaling have been used in the analysis
of slow modes within resonant layers in a magnetically structured plasma
(e.g., Ballai et al. 1998) and to study slow MHD waves in a
stratified medium (e.g.,
Roberts 2006, except that in his study the stretching was only on the position
and time since the magnetic field was considered constant).

Considering the thin tube approximation, $\epsilon \ll 1$, i.e., the
$r$-coordinate has a small range in comparison with the $z$-coordinate.
Since we are interested in the longitudinal component of the wave, we focus on
the equation of $\upsilon_z$, and after the scaling it becomes
\begin{eqnarray}
\rho_{0} \left[ \frac{\partial^2 \upsilon_{z}}{\partial \tau^2}-c_\mathrm{T}^2
\frac{\partial^2 \upsilon_{z}}{\partial Z^2} \right] -\frac{c_{\mathrm s}^{2}}
{c_{\mathrm f}^{2}} \frac{B_R^2}{\mu} \frac{\partial^2 \upsilon_{z}}{\partial R^2} 
-\frac{c_{\mathrm s}^{2}}{c_{\mathrm f}^{2}} \frac{2 B_R B_z}{\mu}
\frac{\partial^2 \upsilon_{z}}{\partial R \partial Z} +\left[ \left(
\frac{c_{\mathrm T}^{2}}{c_{\mathrm f}^{2}}-\frac{c_{\mathrm s}^{4}}{c_{\mathrm f}^{4}} \right)
\frac{B_z}{\mu} \frac{\partial B_z}{\partial Z}\right] \frac{\partial
\upsilon_z}{\partial Z} \nonumber \\
+\left[ \frac{c_{\mathrm T}^{2}}{c_{\mathrm f}^{2}} \frac{2 B_R}{\mu}
\frac{\partial B_z}{\partial Z}
-\frac{c_{\mathrm s}^{2}}{c_{\mathrm f}^{2}} \left( \frac{B_R}{\mu}
\frac{\partial B_R}{\partial R}
+ \frac{B_z}{\mu} \frac{\partial B_R}{\partial Z} \right) \right]
\frac{\partial \upsilon_z}{\partial R} 
+\left[ \frac{c_{\mathrm s}^{2}}{c_{\mathrm f}^{2}} \frac{B_z}{\mu}
\frac{\partial^2 B_{z}}{\partial Z^2} + \left( \frac{c_{\mathrm s}^{4}}
{c_{\mathrm f}^{4}}-\frac{c_{\mathrm T}^{2}}{c_{\mathrm f}^{2}} \right)
\frac{1}{\mu} \left( \frac{\partial B_z}
{\partial Z}\right)^2 \right] \upsilon_z = 0, \label{eq18}
\end{eqnarray}
where $c_{\mathrm A}= {(B_z^2 /\mu \rho_0)}^{1/2}$ and
$c_{\mathrm s}={(\gamma p_0 / \rho_0)}^{1/2}$ are the Alfv\'{e}n and
sound speeds, respectively. The square of the fast phase speed is defined by
$c_{\mathrm f}^{2}=c_{\mathrm s}^2 + c_{\mathrm A}^2$ and the tube speed is given
by $c_{\mathrm T}^{-2} = c_{\mathrm s}^{-2} + c_{\mathrm A}^{-2}$.

The $R$-derivatives of perturbed quantities are very small compared with
the $Z$-derivatives since we consider the thin tube approximation, i.e.,
$\partial \upsilon_z /\partial R \ll 1.$

\section{Pressure wave equation}
\label{pressure}
In this Appendix, we present the governing pressure equation and the relationship between velocity and pressure amplitude for longitudinal stratification.
Equations (\ref{eq07})-(\ref{eq13}) may also be combined to give an
equation for the (perturbation) pressure amplitude $p(z)$, defined by
$p(z,t)=p(z) \exp (i\omega t)$.  If one follows this route, one will arrive at
\begin{equation}
\frac{d^2 p}{d Z^2} - \left( \frac{1}{\rho_0}\frac{d \rho_0}{d Z}
+\frac{1}{B_z} \frac{d B_z}{d Z} \right) \frac{d p}{d Z} +
\frac{\omega^2}{c_{\mathrm T}^2} p=0, \label{eq22}
\end{equation}
after applying the same scaling as defined by (\ref{eq17}). Equation (\ref{eq22})
is consistent with the results obtained by Roberts \& Webb (1978).
From Equation (\ref{eq08}), the following relation is obtained:
\begin{equation}
\frac{\partial p}{\partial Z}= -\rho_0 (Z) \frac{\partial \upsilon_z}
{\partial \tau}. \label{eq23}
\end{equation}
We can find the velocity amplitude equation from the pressure equation
by using (\ref{eq23}):
\begin{eqnarray}
\frac{d^2 \upsilon_{z}}{d Z^2}+ \left(\frac{1}{c_{\mathrm T}^2}
\frac{\partial c_{\mathrm T}^2}{\partial Z}+\frac{1}{\rho_0}
\frac{\partial \rho_0}{\partial Z}-\frac{1}{B_z}
\frac{\partial B_z}{\partial Z} \right) \frac{d \upsilon_z}{d Z} 
+\left[ \frac{\omega^2}{c_{\mathrm T}^2}
-\frac{1}{B_z} \frac{\partial^2 B_{z}}{\partial Z^2}
+ \frac{1}{B_z^2} \left( \frac{\partial B_z}{\partial Z}\right)^2 \right]
\upsilon_z \nonumber \\
+\left[ -\frac{1}{\rho_0} \frac{\partial \rho_0}{\partial Z}
\frac{1}{B_z} \frac{\partial B_z}{\partial Z}-
\frac{1}{c_{\mathrm T}^2}\frac{\partial c_{\mathrm T}^2}{\partial Z}
\frac{1}{B_z} \frac{\partial B_{z}}{\partial Z}
\right] \upsilon_z = 0, \label{eq24}
\end{eqnarray}
which is indeed equivalent to Equation (\ref{eq19}) and that can be
checked rather easily.


\end{document}